\documentstyle[twocolumn,aps,epsfig,floats]{revtex}
\begin{document}

\twocolumn[\hsize\textwidth\columnwidth\hsize\csname %
@twocolumnfalse\endcsname

\title{One-dimensional Surface Bound States in d-wave Superconductors}
\author{Dirk K. Morr $^1$ and Eugene  Demler $^2$ }
\address{$^1$ Theoretical Division, Los Alamos National
Laboratory, Los Alamos, NM 87545 \\ $^2$ Department of Physics,
Harvard University, Cambridge, MA 02138}
\date{\today}
\draft \maketitle

\begin{abstract}
We present an {\it exact} quantum theory for the
bound states in the vicinity of an
edge or a line of impurities in a $d_{x^2-y^2}$ superconductor.
For a $(110)$-surface we show that a finite dispersion
of the one dimensional band of bound states leads to a  two peak
structure in the density of states (DOS). We study the
effect of an applied magnetic field and a subdominant $id_{xy}$
order parameter on the DOS and discuss the implications of our results for
tunneling experiments.
\end{abstract}


]

The existence of a zero bias conductance peak (ZBCP), observed by
in-plane tunneling spectroscopy on YBa$_2$Cu$_3$O$_7$ (YBCO) thin films
\cite{Lesueur,Geerk,Covington0,Alff} has been interpreted as a
signature of unconventional superconductivity.
In the framework of the quasiclassical approximation, it was
first shown by Hu \cite{Hu}, and subsequently studied in more details in
Refs.~\cite{Kashiwaya,Fogelstrom}, that
in superconductors with $d_{x^2-y^2}$ symmetry the ZBCP arises
from a one-dimensional dispersionless band of surface bound states.
Recently Covington {\it et al.}~discovered the splitting of the ZBCP into two
peaks not only in the presence of an applied field, $H$, but also for $H=0$
below $T_s=8$ K \cite{Covington}. Within the quasiclassical approach, the
splitting for $H=0$ was ascribed to a spontaneously broken time
reversal symmetry (BTRS) on the surface of YBCO and
the generation of a subleading superconducting order parameter
with $i d_{xy}$ or $i s$ symmetry \cite{Fogelstrom}. While in conventional
superconductors corrections to the bound state energy beyond the quasiclassical
approximation are too small to be observed experimentally, the corrections in
high temperature superconductors (HTSC) are of the order of
$\Delta^2_0/E_F \approx 3$ meV and thus comparable to the
experimentally observed peak splitting. This necessitates a fully
microscopic treatment of the problem before a conclusion about
the nature of the splitting for $H=0$ can be reached.

In this Letter we present a microscopic theory which goes
beyond the quasiclassical approximation and allows us to capture
properties of the ZBCP that eluded earlier investigations.
By developing a scattering matrix formalism for one-dimensional (1D)
defects, appropriate to describe surfaces or lines of impurities,
we show that the band of bound states has a finite dispersion
that leads to a two-peak structure in the
density of states (DOS), even in the absence of a BTRS.
We predict the existence of Friedel oscillations associated with
the bound states and show that their spatial dependence varies
strongly with the so-called acceptance angle of the measurements.
In the presence of a supercurrent or magnetic field the DOS exhibits a
four-peak structure for small acceptance angles, but only two
peaks with increased frequency splitting for large acceptance
angles. We show that the onset of a subdominant order parameter with
$id_{xy}$ symmetry leads  to a sharpening of the two-peak structure in
the DOS, but does not induce any qualitatively new features.
Finally, we compare our results with the in-plane tunneling
experiments on YBCO \cite{Covington} and comment on
the extensions of our formalism to treat  surface bound states in
systems with other non-trivial order parameters.

We begin by presenting an {\it exact} scattering theory which
describes the emergence of a bound state in the vicinity of an
interface, e.g., a line of impurities or a surface along the
$(1m0)$ direction, in a $d_{x^2-y^2}$ superconductor. The
scattering of the electrons off the interface is described by
$U({\bf r})= U \sum_{{\bf R}_i} \delta({\bf r}-{\bf R}_i)$,
where the sum runs over all ${\bf R}_i=(x,m x)$.
The Green's function in the presence of this potential is given
by
\begin{eqnarray}
\hat{G}({\bf r},{\bf r}',\omega) &=& \hat{G}_0({\bf r}-{\bf
r},\omega) + \nonumber \\ & & \hspace{-1.0cm} \sum_{{\bf s} \,
{\bf s}'}  \hat{G}_0({\bf r}-{\bf s},\omega) \hat{T}({\bf
s},{\bf s}',\omega) \hat{G}_0({\bf s}'-{\bf r}',\omega) \ ,
\label{fullG}
\end{eqnarray}
where the fourier transform of the scattering $\hat{T}$-matrix only depends on
the momentum
parallel to the line, $q_\parallel$,
\begin{equation}
\hat{T}(q_\parallel,\omega) = [ 1-U \hat{\tau}_3  \int^\prime
 { d {\bf p} \over 2 \pi } \, \hat{G}_0({\bf p},\omega) ]^{-1} \,\,
U \hat{\tau}_3 \ .
\label{Tmatrix}
\end{equation}
$\hat{\tau}_{\alpha}$ are the Pauli matrices, and the prime restricts the
integration to those momenta in the first Brillouin zone (BZ) for which $p_x+m
p_y-q_\parallel=2 \pi n$, with $n$ integer. For the Green's function of the
clean system one has
\begin{equation}
\hat{G}_0({\bf k},\omega)=
[\omega \hat{\tau}_0- \epsilon_{\bf k}
\hat{\tau}_3 - \Delta^{(0)}_{\bf k} \hat{\tau}_1]^{-1} \ ,
\end{equation}
where $\Delta^{(0)}_{\bf k}=\Delta_0(\cos k_x - \cos k_y)/2$ is
the $d_{x^2-y^2}$ superconducting gap \cite{com2}. For the normal
state we use a quasi-particle dispersion, representative for the
HTSC
\begin{equation}
\epsilon_{\bf k}=- 2 t ( \cos k_x+ \cos k_y)
-4 t' \cos k_x \cos k_y - \mu
\end{equation}
with $t=300$ meV, $t'/t=-0.3$, $\mu/t=-0.99$  and $\Delta_0=25$ meV \cite{com1}.

In the following we consider for definiteness a surface along the
$(110)$ direction (with $U=\infty$). The induced surface bound
states represent quasiparticles that are localized in the
direction perpendicular to the surface, but have a well defined momentum
parallel to it. The dispersion, $\omega_{s}({q_\parallel})$, of these bound
states is determined by the poles of the $\hat{T}$-matrix
and thus follows from the condition
\begin{equation}
\int^\prime d {\bf p} \,  { \epsilon_p \over
\omega^2_{s}-\epsilon_p^2-\Delta_p^2 } =
 \pm  \int^\prime  d {\bf p} \,
{ \omega_{s} \  \over
\omega^2_{s}-\epsilon_p^2-\Delta_p^2} \ . \label{tp}
\end{equation}
For a $(110)$-surface, the integration is restricted to a line of momenta
parallel to the $(1{\bar 1} 0)$-direction (dashed line in the inset of
Fig.~\ref{poles}).
\begin{figure} [t]
\begin{center}
\leavevmode
\epsfxsize=7.5cm
\epsffile{Fig1.ai}
\end{center}
\caption{Dispersion, $\omega_s(q_\parallel)$, (solid line) and gap edge of the
particle-hole continuum (dashed line) of the surface bound
states as a function of $q_\parallel$. Inset:  characteristic FS of the
HTSC and lines of integration for a $(110)$-surface. We set the lattice constant
$a_0=1$.} \label{poles}
\end{figure}
For small $q_\parallel$ the integrals in Eq.(\ref{tp}) can be
expanded by standard means, and one finds that the dispersion
is quadratic in $q_\parallel$ if the Fermi surface (FS)
is particle-hole symmetric and linear in $q_\parallel$ when this symmetry
is absent (as is the case for the FS shown in Fig.~\ref{poles}).
In Fig.~\ref{poles} we present the dispersion of the bound states,
obtained from the full numerical solution of Eq.(\ref{tp}), together with
the gap edge of the particle-hole continuum.  The
sharp drop of the bound state energy at $q_\parallel^0$ occurs
when the line of integration crosses the FS at ${\bf k}_c$ (see
inset). Just below this crossing, i.e., for $q_\parallel \lesssim
q_\parallel^0$, the poles of the $\hat{T}$-matrix lie within the
particle-hole continuum, and the surface states are damped
resonant state.

The DOS, renormalized by the defect/surface scattering, at a distance
$r_{\perp}=(r_x-r_y)/\sqrt{2}$ from the surface is given by
$
N(\omega,r_{\perp}) = {\rm Im} \, [  \hat{G}_{0}({\bf r},{\bf
r},\omega) + \delta   \hat{G}({\bf r},{\bf r},\omega) ]_{\,11}$,
where
\begin{eqnarray}
\delta   \hat{G}(r_\perp,\omega)
&=&\int_{-\Lambda}^{\Lambda} \frac{dq_\parallel}{2\pi}
\hat{G}^{\prime \prime}_0(\omega,q_\parallel, {\bf r})
\hat{T}(q_\parallel,\omega)
\hat{G}^\prime_0(\omega,q_\parallel, - {\bf r}) \nonumber \\
& & \hspace{-.5cm} \hat{G}^{\prime (\prime \prime)}_0(\omega,q_\parallel,
{\bf r}) =\int^{\prime (\prime \prime)} { d {\bf p} \over 2 \pi} \,
e^{i {\bf p} {\bf r}} \hat{G}_0(\omega,{\bf p}) \ ,
\label{dG}
\end{eqnarray}
and the double prime restricts the integration to all momenta in
the first BZ with $p_x+m p_y-q_\parallel=0$. Summation over all
$q_\parallel$ in Eq.(\ref{dG}) corresponds to $\Lambda=
\sqrt{2}\pi$. However, in the analysis of tunneling experiments it
is in general assumed that $\Lambda < \sqrt{2} \pi$ due to the
finite acceptance angle of the experiments \cite{Fogelstrom}; in
what follows we therefore consider two  different values of
$\Lambda$ in our calculations of the DOS. Note that the above
expression for $\delta \hat{G}$ is similar to the one for a single
impurity problem \cite{Shiba}.

In Fig.~\ref{DOS} we plot $N(r_\perp,\omega)$ obtained numerically
for $\Lambda=1/\sqrt{2}$ and $\sqrt{2}\pi$.
\begin{figure} [t]
\begin{center}
\leavevmode
\epsfxsize=7.9cm
\epsffile{Fig2.ai}
\end{center}
\caption{ $N(r_\perp,\omega)$ for {\it (a)} $\Lambda=1/\sqrt{2}$,
and {\it (b)} $\Lambda=\sqrt{2}\pi$. The coloring indicates
the magnitude of $N(r_\perp,\omega)$.}
\label{DOS}
\end{figure}
In both cases, the DOS exhibits a two-peak structure
and Friedel oscillations (FO) whose spatial dependence is
described by $N(r_\perp,\omega) \sim e^{ik_0r_\perp} \
e^{-r_\perp/\xi}$. Both $k_0$ and $\xi$ vary with frequency and
through $\omega_s(q_\parallel)$ implicitly depend on
$q_\parallel$. For $\Lambda=1/\sqrt{2}$, there exist only {\it
two} bound states with wave-vector $\pm q_\parallel$ for any
frequency $\omega \leq \omega_s(\Lambda)$ and one has
$k_0(\omega_s)=2k_{F,\perp}(\omega_s)=\sqrt{2}(k_F^x-k_F^y)$
(see Fig.~\ref{poles}). Moreover, since $k_0$ increases with increasing
$|q_\parallel|$, the wavelength of the FO decreases with increasing $|\omega|$,
as can clearly be seen in Fig.~\ref{DOS}a. The lengthscale, $\xi$, for the
exponential decay of the FO into the bulk is set by
$\xi=v_{F,\perp}/\sqrt{\bar{\Delta}^2-\omega^2_s} \gg 1$, where
$\bar{\Delta}$ is the superconducting gap at the respective FS
crossing. For $\Lambda=\sqrt{2} \pi$, where we integrate over all
 $q_\parallel$, the FO below and above 4 meV
are {\it qualitatively} different. For $\omega \lesssim 4$ meV
there exist four  bound states, two close to $q_\parallel=0$ and
$q_\parallel=\pi$ (see Fig.~\ref{poles}), whose
superposition leads to the rapid decay of the FO at low
frequencies. In contrast, for $\omega \geq 4$ meV,
only two bound states exist for a given frequency and the DOS again exhibits
well defined FO. While a particle-hole symmetry is absent in the DOS for a given
$r_\perp$, it is recovered
after averaging over $r_{\perp}$.

We next study the effects of an applied magnetic field on the DOS. To avoid the
problems arising from the presence of vortices and a spatial variation of the
supercurrent, we consider for simplicity an average supercurrent
with momentum ${\bf p}_s$; our results are thus exact for the case when a
uniform supercurrent is applied to the sample. For ${\bf p}_s \not = 0$ the
electronic spectrum experiences a Doppler-shift \cite{Tink80}
$
E_{\bf k} = \sqrt{  \epsilon_{\bf k}^2 + |\Delta_{\bf k}|^2} +
{\bf v}_F({\bf k})\cdot{\bf p}_s
$,
where ${\bf v}_F({\bf k})=\partial \epsilon_{\bf k} /\partial {\bf
k}$. Thus, by making the substitution
$
\omega \rightarrow \omega - {\bf v}_F({\bf k})\cdot{\bf p}_s
$
in Eq.(\ref{fullG}), we obtain $N(\omega,r_\perp)$
which we plot in Fig.~\ref{DOSps} for a supercurrent along
the $(110)$-direction with $p_s=0.015$.
\begin{figure} [t]
\begin{center}
\leavevmode
\epsfxsize=7.5cm
\epsffile{Fig3.ai}
\end{center}
\caption{$N(\omega,r_\perp)$ for $p_s=0.015$ and {\it (a)}
$\Lambda=1/\sqrt{2}$, and {\it (b)} $\Lambda=\sqrt{2}\pi$ }
\label{DOSps}
\end{figure}
The bound state energy is now shifted to
$\bar{\omega}_s(q_\parallel)= \omega_s(q_\parallel) + {\bf v}_F
\cdot{\bf p}_s$, where for $\pm q_\parallel$, ${\bf v}_F$ is the
Fermi velocity at the respective FS crossing. This shift leads to
a four-peak structure in the DOS, as can be seen in Fig.~3a for
$\Lambda=1/\sqrt{2}$. Since the energy shift depends on
$q_\parallel$ through ${\bf v}_F$, the peaks which are shifted
towards $\omega=0$ are sharpened, but still possess the same
spectral weight as the apparently weaker peaks shifted to larger
$|\omega|$. In contrast, for $\Lambda=\sqrt{2} \pi$
(Fig.~\ref{DOSps}b), the Doppler-shift varies much more strongly
along the lines of integration which preserves the two-peak
structure in the DOS, increases the frequency splitting between
the two peaks, and leads to a filling-in of the gap between the
peaks. Thus, we predict that only for small acceptance angles can
a four-peak structure be resolved.

We now turn to the effect of a  subdominant order parameter with
$d_{xy}$ symmetry and gap $\Delta^{(1)}_{\bf k}=\Delta_1 \sin k_x
\, \sin k_y$ on the DOS. In Fig.~\ref{comp} we plot
$N(\omega,r_\perp)$ for a superconducting gap with $d_{x^2-y^2}$
(solid line) and $d_{x^2-y^2}+id_{xy}$ symmetry (dashed line). For
the latter, we assume that the total superconducting gap is given
by $\Delta^{(0)}_{\bf k}+i\Delta^{(1)}_{\bf k}$ with
$\Delta_1/\Delta_0=0.1$ \cite{com3}.
\begin{figure} [t]
\begin{center}
\leavevmode
\epsfxsize=7.5cm
\epsffile{Fig4.ai}
\end{center}
\caption{$N(\omega,r_\perp)$ as a function of $\omega$ for
$\sqrt{2}r_\perp=1.25$ for a a superconducting gap with
$d_{x^2-y^2}$ (solid line) and $d_{x^2-y^2}+id_{xy}$ symmetry
(dashed line).}
\label{comp}
\end{figure}
The DOS at small $\omega$ increases linearly with energy for the $d_{x^2-y^2}$
case and exhibits as expected a gap for the $d_{x^2-y^2}+id_{xy}$ case. Note,
that the the onset of an $id_{xy}$ component shifts the two peaks to only
slightly higher frequencies.

We next comment on the differences between our scattering matrix
approach and earlier quasiclassical approaches within the
Bogolyubov-deGennes formalism.  In the latter, one considers only
quasiparticles with a well defined momentum on the Fermi surface,
such that for a $(110)$-surface the incident and reflected
quasiparticles experience a superconducting gap of equal magnitude
and opposite sign, which leads to a zero energy bound states for
all $q_\parallel$ \cite{Hu,Kashiwaya,Fogelstrom}. This
result is exact for $q_\parallel= 0$, since here $\Delta({\bf k})
\equiv 0$, but for any finite $q_\parallel$, the scattering
between states away from the FS introduces corrections to the
quasiclassical energy. Hence, within our scattering matrix
formalism, the quasiclassical results are recovered by restricting
the integration in Eq.(\ref{tp}) to the Fermi surface which yields
$\omega_s=0$ for {\it all} $q_\parallel$. The appearance of a
dispersing band of bound states is thus a pure {\it quantum}
effect \cite{Wal97}, which cannot be obtained by a quasiclassical
approximation.

Experiments of ~\cite{Covington} on YBCO materials show a single
peak above $T_s=8\ {\rm K} \ll T_c$, and the emergence of a
two-peak structure at lower temperatures. This and the small
increase in the overall width of the ZBCP was attributed to the
appearance of a subleading order parameter ($id_{xy}$ or $is$)
below $T_s$. However, the observed broadening of the ZBCP below
$T_s$ is much smaller than the splitting between the peaks. This
is expected if the peak splitting arises not from the appearance
of a BTRS, but from the finite bound state dispersion discussed
above. A subleading order parameter may sharpen the two-peak
structure at lower temperatures and suppress the low energy part
of the spectrum but is not predominantly responsible for the peak
splitting. Moreover, recent experiments on
Bi$_2$Sr$_2$Ca$_2$O$_{8+\delta}$ \cite{Greene2}, which find that
the width of the ZBCP does not increase with the emergence of a
two-peak structure in the DOS, are difficult to explain within the
BTRS scenario. Currently, experiments using electron-spin
resonance \cite{Pugel} and phase sensitive measurements
\cite{Dale} are under way to determine whether a BTRS exists on
the surface of HTSC. In addition, detailed measurements of the low
energy part of the DOS also provide important information since
the DOS for a $d_{x^2-y^2}$ superconductor is linear at low
frequencies, whereas that for $d+id$ or $d+is$ order parameters is
fully gapped. Another interesting feature of the experiments
\cite{Covington} is that only two peaks are observed in an applied
magnetic field, which shift to higher energies with increasing
field. One possible explanation of this result is that the
acceptance angle of the experiment (see Fig.~\ref{DOSps}b) is
large, in which case the regime of smaller acceptance angles can
be accessed in tunneling experiments by using stronger barriers
between the superconducting and normal materials. In this limit,
we predict that an applied magnetic field induces a four-peak
structure in the DOS (Fig.~\ref{DOSps}a). An alternative
explanation is that the acceptance angle is small but that the
magnetic field generates a large $id$ (or $is$) component at the
boundary that pushes the peaks to higher energies
(Fig.~\ref{comp}). Each of the original peaks should then be split
into two, however such splitting may be too small to be observed
experimentally.  The latter scenario is consistent with a strong
decrease in the DOS at zero energy with increasing magnetic field
\cite{Covington}.

Finally, the bound states discussed above provide an interesting
new class of one dimensional fermions whose dispersion can be
changed continuously by applying a magnetic field (or
supercurrent), with the possibility of making all fermions chiral
\cite{unpublished}. Moreover, the formalism presented above can be
generalized to study surface bound states in systems with other
non-trivial order parameters, such as $d$-density or spin density
wave states suggested recently \cite{ddw}. We find that a zero
energy bound states also exists for a $d$-density wave state in
the vicinity of a $(110)$ surface. This bound state can be
observed in tunneling experiments and may thus prove to be an
important feature of the underdoped cuprates; an account of our
results will appear elsewhere \cite{unpublished}.

In summary we present an {\it exact} scattering theory for bound
states which are induced close to lines of impurities or surfaces
in a $d_{x^2-y^2}$ superconductors. We find that these bound
states disperse linearly with momentum $q_\parallel$, a dispersion
that gives rise to a two-peak structure in the density of states.
For small acceptance angles, we predict that an applied magnetic
field leads to a splitting of the two peaks into four, while for a
large acceptance angle, there exist only two peaks. Our results
are consistent with recent in-plane tunneling experiments on YBCO
and provide important insight into their interpretation. We
present detailed predictions for the spatial dependence of the
Friedel oscillations in the DOS that may be measured by the STM
experiments near the surfaces, grain boundaries, or lines of
impurities.

We  thank L. H. Greene, D. Van Harlingen, and L. Marinelli for
stimulating discussions. DM acknowledges support by the Department
of Energy. ED is supported by the Harvard Society of Fellows and
the Milton Fund.


\begin{thebibliography}{99}
\vspace{-12pt}


\bibitem{Lesueur} J. Lesueur {\it et al.}, Physica {\bf C 191}, 325 (1992).

\bibitem{Geerk} J. Geerk {\it et al.}, Z. Phys. {\bf 73}, 329 (1994).

\bibitem{Covington0} M. Covington {\it et al.}, Appl. Phys. Lett. {\bf
68}, 1717 (1996).

\bibitem{Alff} L. Alff {\it et al.}, Phys. Rev. B {\bf 55}, R14757 (1997).

\bibitem{Hu} C.R. Hu,  Phys. Rev. Lett. {\bf 72}, 1526 (1994).

\bibitem{Kashiwaya}
S. Kashiwaya, Y. Tanaka, M. Koyanagi, and K. Kajimura, Jpn. J.
Appl. Phys {\bf 34}, 4555 (1995); Yu. S. Barash, A.A.
Svidzinsky, and H. Burkhardt, Phys. Rev. B {\bf 55}, 15282 (1997)

\bibitem{Fogelstrom} M. Fogelstr\"{o}m, D. Rainer, and J.A. Sauls,
Phys. Rev. Lett. {\bf 79}, 281 (1997); D. Rainer {\it et al.},
cond-mat/9712234; S. Kos, cond-mat/0008410.

\bibitem{Covington} M. Covington {\it et al.}, Phys. Rev. Lett. {\bf
79}, 277 (1997); M. Aprili, E. Badica, and L.H. Greene, Phys. Rev.
Lett. {\bf 83}, 4630 (1999); L.H. Greene {\it et al.}, Physica B
{\bf 280}, 159 (2000); R. Krupke and G. Deutscher, Phys. Rev. Lett. 83, 4634-7
(1999).

\bibitem{com2} We neglect the suppression of $\Delta(k)$
close to the surface since it does not change our results
qualitatively.

\bibitem{com1} Note that our results are qualitatively robust
against changes in the electronic band structure.

\bibitem{Shiba} H. Shiba, Prog. Theor. Phys. {\bf 40}, 435 (1968).

\bibitem{Tink80} {\it Introduction to Superconductivity} by M.
Tinkham, Krieger Publishing, 1980.

\bibitem{com3} Appreciable corrections to our results are only expected if the
$id_{xy}$ component decays away from the surface on a length scale which is
considerably
shorter than the width of the bound state.

\bibitem{Wal97} M. B. Walker, P. Pairor, and M. E. Zhitomirsky,
Phys. Rev. B {\bf 56}, 9015 (1997).

\bibitem{Greene2} H. Aubin {\it et al.}, to appear in Physica C.

\bibitem{Pugel} D.E. Pugel {\it et al.}, preprint.

\bibitem{Dale} W.K. Neils and D. J. Van Harlingen,
Physica B {\bf 284}, 587 (2000); W.K. Neils,  B.L.T. Plourde and D.J. Van
Harlingen,
Physica C {\bf 341-348}, 1705 (2000).

\bibitem{unpublished} E.A. Demler and D.K. Morr, unpublished.

\bibitem{ddw} C. Nayak, cond-mat/0001303; S. Charavarty {\it et.al},
cond-mat/0005443.

\end{thebibliography}
\end{document}